\RequirePackage{fix-cm}
\documentclass[smallextended]{svjour3}       
\smartqed  
\usepackage[colorlinks=true, citecolor={blue}, urlcolor={black}]{hyperref}
\usepackage{graphicx}
\usepackage{amsmath}
\usepackage{amssymb}
\usepackage{color}
\usepackage{caption}
\usepackage{latexsym}

\begin{document}

\title{Modeling of rigidity dependent CORSIKA simulations for GRAPES-3
\thanks{$^\ast$Corresponding authors}}

\author{B. Hariharan        \and
        S.R. Dugad          \and
        S.K. Gupta          \and
        Y. Hayashi          \and
        S.S.R. Inbanathan   \and
        P. Jagadeesan       \and
        A. Jain             \and
        S. Kawakami         \and
        P.K. Mohanty        \and
        B.S. Rao
}

\institute{
        B. Hariharan$^\ast$ \and S.R. Dugad \and S.K. Gupta \and P. Jagadeesan 
        \and A. Jain \and P.K. Mohanty \and B.S. Rao \at
            Tata Institute of Fundamental Research, Dr Homi Bhabha Road, Mumbai
            400005, India \\   
            \email{89hariharan@gmail.com} 
            \and
        B. Hariharan$^\ast$ \and S.R. Dugad \and S.K. Gupta \and P. Jagadeesan 
        \and A. Jain \and P.K. Mohanty \and B.S. Rao \and Y. Hayashi \and 
        S. Kawakami \at
            The GRAPES-3 Experiment, Cosmic Ray Laboratory, Raj Bhavan, Ooty
            643001, India
            \and
        B. Hariharan$^\ast$ \and S.S.R. Inbanathan$^\ast$ \at
            The American College, Madurai 625002, India \\
            \email{ssrinbanathan@gmail.com}
            \and
        Y. Hayashi \and S. Kawakami \at
            Graduate School of Science, Osaka City University, 558-8585 Osaka,
            Japan
}

\date{Received: date / Accepted: date}
\maketitle

\begin{abstract}
The GRAPES-3 muon telescope located in Ooty, India records 4$\times$10$^9$ muons
daily. These muons are produced by interaction of primary cosmic rays (PCRs) in
the atmosphere. The high statistics of muons enables GRAPES-3 to make precise
measurement of various sun-induced phenomenon including coronal mass ejections
(CME), Forbush decreases, geomagnetic storms (GMS) and atmosphere acceleration
during the overhead passage of thunderclouds. However, the understanding and
interpretation of observed data requires Monte Carlo (MC) simulation of PCRs and
subsequent development of showers in the atmosphere. CORSIKA is a standard MC
simulation code widely used for this purpose. However, these simulations are
time consuming as large number of interactions and decays need to be taken into
account at various stages of shower development from top of the atmosphere down
to ground level. Therefore, computing resources become an important
consideration particularly when billion of PCRs need to be simulated to match
the high statistical accuracy of the data.  During the GRAPES-3 simulations, it
was observed that over 60$\%$ of simulated events don't really reach the Earth's
atmosphere. The geomagnetic field (GMF) creates a threshold to PCRs called
cutoff rigidity R$_c$, a direction dependent parameter below which PCRs can't
reach the Earth's atmosphere.  However, in CORSIKA there is no provision to set
a direction dependent threshold. We have devised an efficient method that has
taken into account of this R$_c$ dependence. A reduction by a factor $\sim$3 in
simulation time and $\sim$2 in output data size was achieved for GRAPES-3
simulations. This has been incorporated in CORSIKA version v75600 onwards.
Detailed implementation of this along the potential benefits are discussed in
this work.

\end{abstract}

\section{Introduction}
PCRs are predominantly the nuclei of hydrogen ($\sim$90\%), helium ($\sim$9\%),
and a small fraction ($\sim$1\%) of remaining heavier elements including carbon,
nitrogen, oxygen, aluminium, iron etc. They span an energy range from sub-GeV to
10$^{11}$\,GeV. The PCR flux steeply decreases with energy exhibiting a power
law spectrum with a spectral slope of -2.7. The bulk of the PCRs ($>$99.99$\%$)
lie below 100\,GeV and are sensitive to inter-planetary magnetic field and solar
wind. Thus, the solar activity modulates the PCR flux at these energies. The PCR
modulation can be used to study both transient solar phenomenon such as CME,
GMS and long-term solar phenomenon related to seasonal, 11-, and
22-year solar cycle over the past eight decades.  

The GMF acts as a shield by deflecting out low energy PCRs.  Below a certain
rigidity value, the PCRs do not reach the Earth's atmosphere. This threshold is
known as the geomagnetic cutoff rigidity (R$_c$).  Since the strength of the GMF
varies over the Earth, the R$_c$ is also strongly dependent on the location on
the Earth. R$_c$ is almost zero at the geomagnetic poles and is about 15\,GV at
the geomagnetic equator.  Although the PCR distribution is known to be almost
isotropic, however, the PCR flux observed at the top of the atmosphere is
anisotropic due to the variation of R$_c$. The PCRs above a few hundred GeV are
least affected by the GMF. 

PCRs after entering into Earth's atmosphere interact with the air nuclei and
produce secondary particles. These secondary particles further interact down in
the atmosphere and produce shower of particles including $\gamma$-rays,
electrons, muons and hadrons at the ground level. This process is called cascade
shower or extensive air shower (EAS). Unlike other particles, muons due to their
energy loss primarily by ionization mostly survive to the ground level. Thus,
the muon flux represents a good proxy of the PCR flux. The flux variation in the
PCR caused due to various phenomenon as mentioned above could be studied through
the measurement of muon flux. In addition, muon flux is also modulated by the
atmospheric parameters such as pressure and temperature
\cite{Mohanty2016_1,Arunbabu2017}. Study of the muon flux variation during
thunderstorms has emerged as an exciting area \cite{Hariharan2019}. 

Interpretation of experimental data requires detailed modeling of EAS
development in the atmosphere. This can be studied with standard MC simulation
packages like CORSIKA \cite{corsika}, CRY \cite{cry}, AIRES \cite{aires}. Since
these codes have to track large number of particles while taking into account of
their hadronic and electro-magnetic interactions, decay processes etc. from the
top of the atmosphere to the ground level, the simulations become CPU intensive.
The simulation time as well as the generated data size increases linearly with
the energy of the PCR. Although the simulation time for low energy PCRs in the
GeV energies is not large, however, it becomes an important consideration when
billions of events are required to be simulated to match the high statistical
accuracy of data. Even computing farms with over thousands of CPU cores have to
be engaged over months to simulate a single physics event
\cite{Hariharan2019,Mohanty2016_2}. Thus, any attempt to reduce the simulation
time as well as output data size is useful.

CORSIKA simulates PCRs selected at random in a given energy range and spectral
slope provided through a control file by the user. The direction of PCRs are
chosen randomly from a range of zenith $\theta$ and azimuthal $\phi$ angles
which have to be set in the same control file.  Normally the minima of the
energy range is usually kept well below the lowest R$_c$ for the experiment's
location. However, as discussed before, the R$_c$ is not same for all the
directions. There was no provision in the earlier versions of CORSIKA to set
minimum direction dependent energy in the control file. Thus, in
post-simulation, if the PCR's rigidity value is below the R$_c$ value of its
direction, it had to be rejected since it can not make to the Earth in reality.
Thus a significant amount of computing resources had to be unnecessarily wasted.
In this work, we discuss a method that avoids simulating events which are below
the R$_c$ of respective direction by modifying the CORSIKA code of version
v74000. The initial development was reported to the cosmic ray community
\cite{Hariharan2015}. By considering its potential benefits, it was recognized
by authors of CORSIKA and was made available to the users from CORSIKA version
v75600 onwards since 2017.  The details of the implementation in CORSIKA with
the potential benefits are discussed in following sections.

\section{GRAPES-3 muon telescope}
The GRAPES-3 experiment consists of a large area (560\,m$^2$) tracking muon
telescope (G3MT) is operating at Ooty in India (11.4$^\circ$N, 76.7$^\circ$E,
2200\,m above mean sea level) in conjunction with an array of 400 plastic
scintillator detectors as a part of EAS experiment \cite{Gupta2005}. The G3MT is
designed to (1) obtain nuclear composition of PCRs, (2) discriminate
$\gamma$-rays from the overwhelming background of charged PCRs, and (3) probe
various solar and atmospheric phenomenon.  The scintillator detectors are placed
in a hexagonal geometry with an inter-detector separation of 8\,m. The total
area covered by the array is 25000\,m$^2$. The EAS array records about
3$\times$10$^6$ events per day in the energy range of 10$^{12}$--10$^{16}$\,eV.

The G3MT uses proportional counter (PRC) as basic detector. Each PRC is a
600\,cm long, 10\,cm$\times$10\,cm cross section mild steel tube with a wall
thickness of 2.3\,mm. A G3MT module consists of 232 PRCs arranged in 4 layers,
with alternate layers placed in mutually orthogonal directions which provides an
area of 35\,m$^2$. Two successive layers of PRCs are separated by 15\,cm thick
concrete. The G3MT permits a two-dimensional reconstruction of muon tracks in
two vertical orthogonal planes. The vertical separation of two layers of PRCs in
the same plane is $\sim$50\,cm which allows the muon track direction to be
measured to an accuracy of $\sim$4$^\circ$. The G3MT permits a two-dimensional
reconstruction of muons in 169 directions with the sky coverage of 2.3\,sr. To
achieve an energy threshold of 1\,GeV for vertical muons, a total thickness
$\sim$550\,g\,cm$^{-2}$ in the form of concrete blocks of 2.4\,m thickness is
used as absorber. The concrete blocks have been arranged in the shape of an
inverted pyramid to achieve an energy threshold of sec($\theta$)\,GeV for muons
incident at zenith angle $\theta$ (with coverage up to 45$^\circ$). Four muon
modules housed in a single hall to form a super-module and four super-modules
constitute the G3MT. It collects 4$\times$10$^9$ muons daily which has been
successfully corrected for efficiency and atmospheric variations
\cite{Mohanty2017,Arunbabu2017}. The muons recorded by G3MT are produced by PCRs
in the energy range of 10$^{10}$--10$^{13}$\,eV.

\section{Computation of R$_c$}

\begin{figure}[t]
\begin{center}
\includegraphics*[width=0.60\textwidth,angle=0,clip]{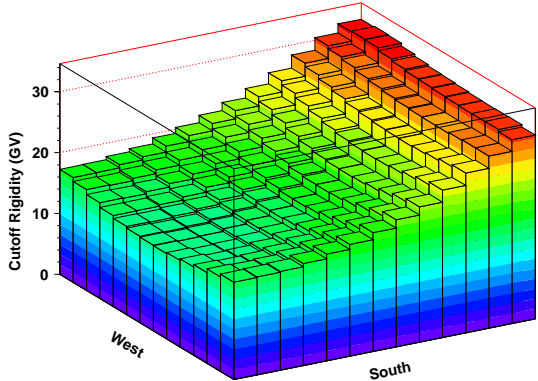}
\vskip -0.05in
\caption{Map of R$_c$ in the field of view (FOV) of G3MT}
\label{fig01}
\end{center}
\vskip -0.35in
\end{figure}

The R$_c$ for a given location can be computed numerically using back-tracing
method \cite{SmartShea}. An anti-proton of a given rigidity is launched from the
observer's location into space and traced in the presence of GMF up to several
Earth radii.  The X, Y, and Z components of GMF are evaluated using IGRF-11
coefficients \cite{IGRF11}. The trajectory calculation is performed for
different rigidities of anti-proton starting from an initial value in decreasing
steps of 0.01\,GV. The rigidity at which the anti-proton trajectory reverses to
the Earth is accepted as R$_c$ of that direction. This means a proton of same
rigidity coming from outside the magnetosphere will be deflected back into
space. R$_c$ is calculated for a grid of 1$^{\circ}\times$1$^{\circ}$ in zenith
$\theta$ and azimuthal $\phi$ angles ranging 0--60$^{\circ}$ and
0--360$^{\circ}$ respectively. Fig.~\ref{fig01} shows a map of R$_c$ for 169
directions in FOV of G3MT, varies in the range of 12--38\,GV.

\section{Implementation of R$_c$ in CORSIKA} 
CORSIKA is a detailed MC simulation package, developed by KIT, Germany to study
the EAS development in the Earth's atmosphere in the energy range of
10$^9$--10$^{20}$\,eV for various primaries. Main CORSIKA code contains about
80000 lines written in FORTRAN, with a few optional subroutines in C++.  It is
interfaced with various external hadronic interaction models such as EPOS,
EPOS-LHC\cite{eposlhc}, QGSJET01C\cite{qgsjet}, QGSJETII-04\cite{qgsII},
SIBYLL\cite{sibyll_2.1}, VENUS\cite{venus}, DPMJET\cite{dpmjet},
NEXUS\cite{nexus} for high energies (calculation of hadron cross-sections above
80\,GeV) and GHEISHA\cite{gheisha}, FLUKA\cite{fluka}, UrQMD\cite{urqmd} for low
energies. An atmospheric model and multiple observational levels can be
specified in the control file by the user. The secondary particles are tracked
till their kinetic energy is above the threshold defined by the user separately
for $\gamma$-rays, electrons, muons, and hadrons. The CORSIKA records physical
quantities like position, momentum and arrival time of secondary particles up to
10 different observational levels.

\begin{figure}[b]
\begin{center}
\includegraphics*[width=0.65\textwidth,angle=0,clip]{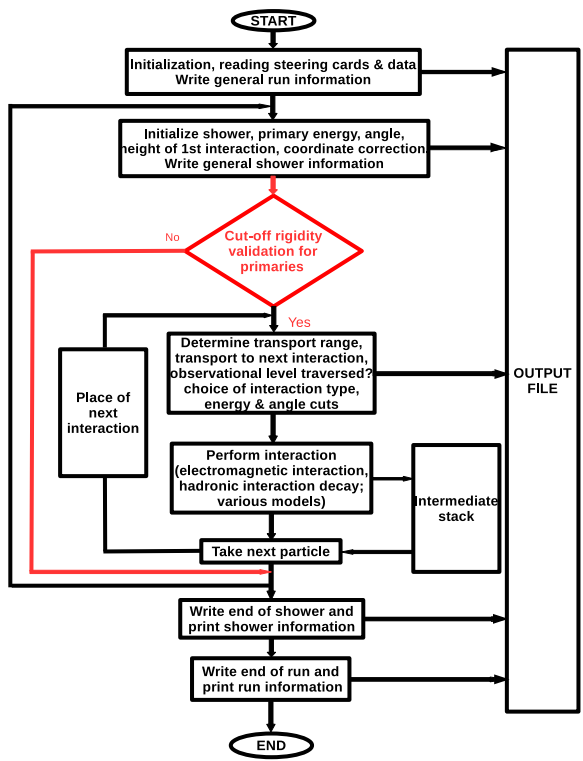}
\vskip -0.05in
\caption{CORSIKA simulation flow (\textcolor[rgb]{1.0,.0,.0}{Modification})}
\label{fig02} 
\end{center}
\vskip -0.35in
\end{figure}

As discussed in the previous section, a database of R$_c$ is generated for the
FOV of G3MT using back-tracing method at a resolution of
1$^{\circ}\times$1$^{\circ}$ in zenith $\theta$ and azimuthal $\phi$ angles
ranging 0--60$^{\circ}$ and 0--360$^{\circ}$ respectively. Since the computation
of R$_c$ is a CPU intensive process, it is difficult to obtain R$_c$ for every
PCR in real-time during simulation. Thus an off-line database is made mandatory
for this implementation and it has to be calculated once by the users for their
observational location.  The coordinate convention used in the back-tracking
program is astronomical standard where $\phi$ moves clockwise from north.  Since
in CORSIKA, $\phi$ moves counter-clock wise from the north, thus 180$^\circ$
rotation is required in database. A simple flow chart of CORSIKA simulation is
displayed in Fig.~\ref{fig02} along with modified flow explaining the R$_c$
validation. During initialization of CORSIKA simulation, the R$_c$ database is
loaded into CORSIKA. Based on the user inputs, energy and direction of a PCR
event is determined at random. Next, a R$_c$ value is determined from the map
based on the direction ($\theta$, $\phi$) of the PCR through a linear
interpolation. If the PCR's rigidity is higher than the R$_c$ of that direction,
then only further simulation of shower development is allowed otherwise it is
rejected. No input and output information of the rejected shower is recorded. As
a result, the simulation time and output data size is reduced significantly.
Since the composition of PCRs are known to be predominantly proton and helium,
the present implementation is made only for proton and helium.  However, this
can be easily extended for other primaries if required.

\section{Results and discussions}
\begin{figure}[t!]
\begin{center}
\includegraphics*[width=0.85\textwidth,angle=0,clip]{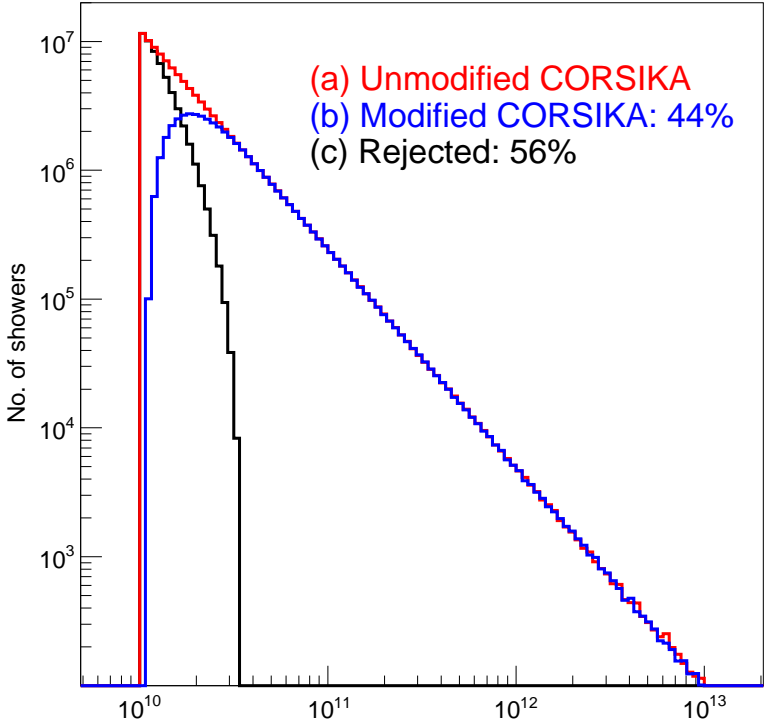}
\vskip -0.05in
\caption{Energy spectrum (eV) of PCR events simulated by (a) unmodified
CORSIKA, and (b) modified CORSIKA. The distribution rejected primaries from
modified CORSIKA is shown in (c).}
\label{fig03}
\end{center}
\vskip -0.35in
\end{figure}

To test the effectiveness of this method, a small set of simulations were
carried out. A total of 10$^8$ proton primaries were generated with unmodified
CORSIKA code over an energy range of 10$^{10}$--10$^{13}$\,eV with a power-law
distribution of spectral slope -2.7 in the angular range of 0--60$^\circ$ and
0--360$^\circ$ for $\theta$ and $\phi$ respectively (hereafter called case A).
The same configuration was repeated with modified CORSIKA code (hereafter called
case B).  In each of these cases, the simulation was carried out by distributing
the total number of events into 1000 jobs in the GRAPES-3 computer cluster
having 1280 CPU cores.  The energy spectrum of simulated events for both the
cases are displayed in Fig.~\ref{fig03}. The spectrum (a) in Fig.~\ref{fig03}
was generated as per given inputs and falls perfectly in a power-law as expected
whereas the spectrum (b) shows only 44\% of input events which underwent the
full shower simulation and the remaining 56\% events which were rejected as they
did not pass the R$_c$ condition of the respective directions are shown in (c).
In case A, by post-simulation processing, 56\% of events were rejected since
they are not really useful for analysis even though they were simulated. In
Fig.~\ref{fig03}, the spectra (a) and (b) are indistinguishable above 35\,GeV.
It is to be noted that R$_c$ validation in CORSIKA are used here with tolerance
of -10\%. The tolerance allows the R$_c$ validation to be checked with lower
R$_c$ of the primary.  This is important for studies where rigidity dependence
is crucial.  The discovery of transient weakening of Earth's magnetic shield
probed by a muon burst is one such example \cite{Mohanty2016_2}. In this study,
the muon burst recorded by G3MT was caused by a G4-class GMS triggered by a CME.
The Earth's magnetic shield was weakened due to magnetic reconnection of GMF
with magnetic field carried by CME. This lowered the R$_c$ in G3MT vicinity
which allowed enormous count of low energy PCRs to enter into Earth's atmosphere
resulted in muon burst for 2 hours. The burst was successfully modeled by MC
simulations which allowed us to use G3MT as rigidity meter. In such studies to
model the rigidity change, it is important to generate CORSIKA events with
tolerance, so that repeated simulations can be avoided.  One would expect higher
rejection ($\sim$63\,\%), leads to larger reduction in simulation time and data
size if 0\% tolerance is given in the validation. However it is recommended to
use with small tolerance, so that the repetition of simulations can be avoided
in future if R$_c$ has to be recalculated or the same data set has to be used
for another EAS experiment. 

\begin{figure}[t]
\begin{center}
\includegraphics*[width=0.85\textwidth,angle=0,clip]{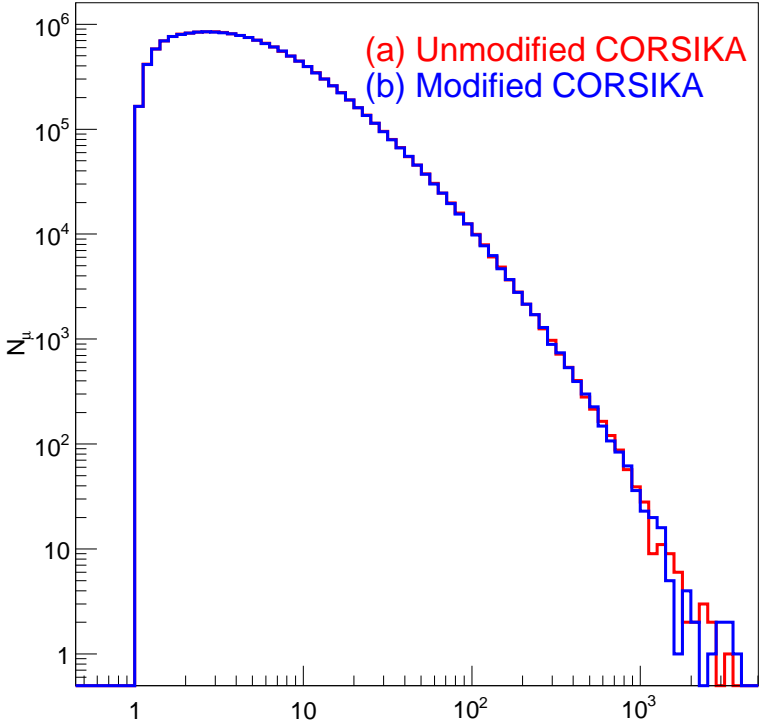}
\vskip -0.05in
\caption{Muon energy spectrum (GeV) obtained from (a) unmodified CORSIKA (b)
modified CORSIKA}
\label{fig04} 
\end{center}
\vskip -0.35in
\end{figure}

The real proof of principle for this implementation is demonstrated by comparing
physics parameters from both these cases. The simulated datasets from both cases
were analysed with in-house developed G3MT detector simulation code. The
in-house simulation code simulates every CORSIKA generated muon in detector
geometry to find hits in four layers of PRCs. It also takes care of angular
threshold and trigger criteria to reconstruct the detected muons in 169
directions using hit patterns which can be easily compared with observation. The
muon energy spectrum for all directions obtained from both cases are compared in
Fig.~\ref{fig04}.  The muon energy spectra obtained from both cases in
Fig.~\ref{fig04} are identical and small perturbations can be seen at higher
energies due to small statistics.  These perturbations found in primary and muon
energy spectra arise due to change in random number sequence in CORSIKA due to
rejected PCR. However, these changes are statistically insignificant.

Table.~\ref{table01} shows quantitative comparison of outputs obtained from both
cases. Significant improvements can be seen in terms of reduction in simulation
time by a factor of $\sim$3 and output data size by a factor of $\sim$2. Despite
of large number of PCRs rejected in case B shown in Fig.~\ref{fig03}, the muon
spectrum from both cases are almost identical as shown in Fig.~\ref{fig04}. This
allows the user to generate more events in disputed time by imposing R$_c$. For
example, a recent result on the measurement of electrical properties for a
thunderstorm event observed in G3MT describes the generation of Giga-Volt
potential in thunderclouds \cite{Hariharan2019}. The thundercloud parameters
were derived with the aid of MC simulations of atmospheric electric field using
CORSIKA. These simulations include generation of 10$^6$ muons for 60 steps of
electric field and 10$^7$ muons for background for each of 169 directions This
extensive simulation was carried out in GRAPES-3 computer cluster running
continuously for two months. It would have taken more than six months and
massive output storage costing $\sim$30\,TB to generate the data bank without
this utility.

One of the biggest threat to our technologies is solar storms which can disrupt
power grids, telecommunications and navigation systems \cite{obama}. Some famous
historic evidences are 'Halloween event', 'Quebec event', and 'Carrington event' 
\cite{Carrington}.  The solar super storm 'Carrington event' recorded in 1859
caused a black out of telegraphic systems in high latitude regions of North
America and Europe. In order to prepare for such solar storms in future, large
number of ground based observatories including network of neutron monitor
stations were installed globally and collecting data throughout the year to
study near Earth phenomenon.  However, the understanding of these phenomenon
primarily depends on precise measurement and accurate MC simulations dealing
large number of PCRs. Because of vast geographical distribution of these
observatories, the R$_c$ varies from 0\,GV at poles to 15\,GV at magnetic
equator. Despite of large number of PCRs were rejected during simulation, the
physics results derived from analysis were unaffected. The present work provides
an advancement in simulating low energy PCRs efficiently by reducing simulation
time and data size. This modification was reported to authors of CORSIKA
which was implemented in official version from v75600 since 2017.

\begin{table}[h!]
\begin{center}
\caption{Comparison of simulation parameters}
\label{table01}
\begin{tabular}{|l|l|l|}
\hline                                                          
\textbf{Parameter} & \textbf{Case A} & \textbf{Case B}\\
\hline                                                          
Number of simulated showers  & $1\times10^8$ & $0.44\times10^8$    \\
Simulation time (min)        & 156           & 46                  \\
File size (GB)               & 239           & 114                 \\
\hline                                                          
\end{tabular}
\end{center}
\end{table}

\begin{acknowledgements}
We thank Dr.~T.~Pierog and Dr.~D.~Heck for implementing this modification in
official CORSIKA releases to benefit cosmic ray community. We thank
D.B.~Arjunan, A.~Chandra, V.~Jeyakumar, S.~Kingston, K.~Manjunath, S.D.Morris,
S.~Murugapandian, P.K.~Nayak, S.~Pandurangan, B.~Rajesh, P.S.~Rakshe,
K.~Ramadass, K.~Ramesh, L.V.~Reddy, V.~Santhoshkumar, M.S.~Shareef, C.~Shobana,
R.~Sureshkumar, and M.~Zuberi for their assistance in running the GRAPES-3
experiment.

\end{acknowledgements}


\end{document}